\documentclass[aps,twocolumn,superscriptaddress,showpacs,longbibliography]{revtex4-1}

\usepackage{graphicx}
\usepackage{dcolumn}
\usepackage{bm}
\usepackage{upgreek}
\usepackage{paralist}
\usepackage{gensymb}
\usepackage{braket}

\newcommand{\EECS}{\affiliation{Research Laboratory of Electronics, Massachusetts Institute of Technology, Cambridge, MA 02139, USA}}

\begin{document}

\title{Two-Dimensional Photonic Crystal Slab Nanocavities on Bulk Single-Crystal Diamond}

\author{Noel H. Wan}
\email{noelwan@mit.edu}
\EECS
\author{Sara Mouradian}
\EECS
\author{Dirk Englund}
\email{englund@mit.edu}
\EECS

\begin{abstract}
Color centers in diamond are promising spin qubits for quantum computing and quantum networking. In photon-mediated entanglement distribution schemes, the efficiency of the optical interface ultimately determines the scalability of such systems. Nano-scale optical cavities coupled to emitters constitute a robust spin-photon interface that can increase spontaneous emission rates and photon extraction efficiencies. In this work, we introduce the fabrication of 2D photonic crystal slab nanocavities with high quality factors and cubic wavelength mode volumes -- directly in bulk diamond. This planar platform offers scalability and considerably expands the toolkit for classical and quantum nanophotonics in diamond.

\end{abstract}
\maketitle
\section*{Introduction}
Defect centers in diamond have emerged as leading atom-like systems for quantum information processing ~\cite{doherty2013nitrogen,childress2013diamond}. Recent demonstrations of spin-photon entanglement with the diamond nitrogen vacancy (NV) center have enabled important steps towards quantum networks, including remote NV-NV spin entanglement~\cite{bernien2013heralded} and entanglement purification of two quantum nodes~\cite{kalb2017entanglement}. However, these processes remain inefficient because of poor photon collection and largely incoherent NV optical transitions, as the zero-phonon line (ZPL) constitutes less than 4\% of the NV's spontaneous emission naturally. A variety of methods have been investigated to improve the NV's ZPL collection efficiency. Solid immersion lenses~\cite{hadden2010strongly,marseglia2011nanofabricated,jamali2014microscopic}, dielectric antennas~\cite{riedel2014low,li2015efficient}, waveguides~\cite{babinec2010diamond,momenzadeh2014nanoengineered,mouradian2015scalable,patel2016efficient}, and parabolic reflectors~\cite{wan2017efficient} address the problem of collection efficiency but do not significantly improve the spontaneous emission fraction into the ZPL. Recently, the negatively-charged silicon-vacancy (SiV) center has gained attention as another promising spin qubit system~\cite{sukachev2017silicon} with a more favorable ZPL emission fraction (Debye-Waller factor) of 0.7 and long-term spectral stability~\cite{becker2017coherence}. However, the SiV's overall radiative quantum efficiency is rather low, at 10-30\%~\cite{riedrich2014deterministic,sipahigil2016integrated}. 

\begin{figure}[ht]
\begin{center}
\includegraphics[width=3.5in]
{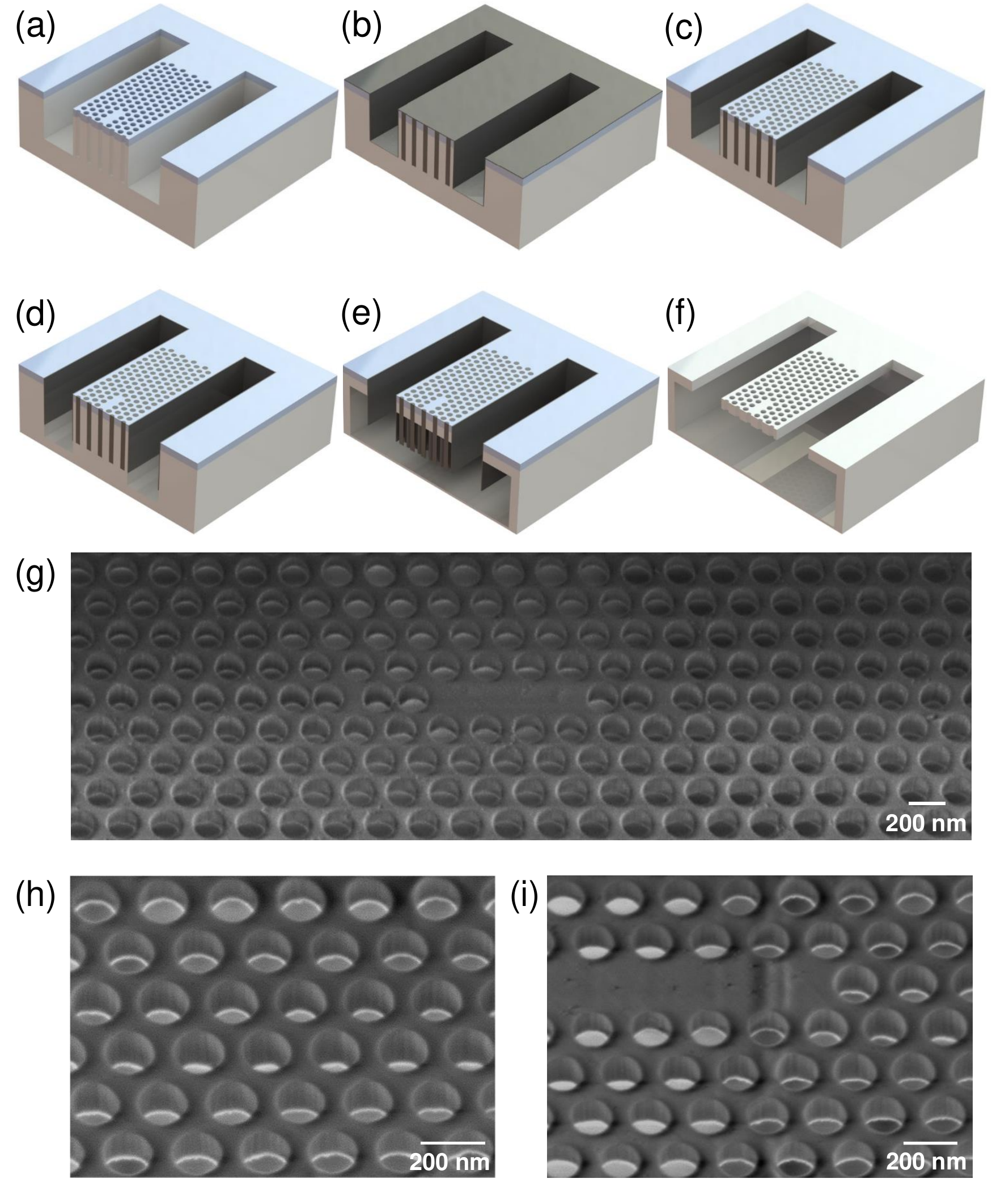}
\end{center}
\caption{
(a) Electron-beam lithography and oxygen plasma reactive-ion etching (RIE) of diamond (grey) using a SiN hard mask (blue). (b) Atomic layer deposition (ALD) of alumina (black) for conformal protection. (c) Break-through etch of alumina using tetrafluoromethane. (d) RIE of diamond. (e) Quasi-isotropic undercut of diamond using oxygen plasma (f) Mask removal using hydrofluoric acid. (g-i) Scanning electron micrograph (SEM) of a 2D photonic crystals suspended in air.
}
\label{fab}
\end{figure}

Photonic nanocavities can improve the efficiency of spin-photon interfaces through spontaneous emission rate modification of cavity-resonant transitions~\cite{purcell1995spontaneous}. In particular, the overall emission rate and fraction into the ZPL increases with the cavity Purcell factor in proportion to the quality factor ($Q$) to mode volume ($V_{\text{m}}$) ratio. A variety of approaches have resulted in high-$Q/V_{\text{m}}$ diamond-based 1D photonic crystal (PhC) nanocavities coupled to isolated NV$^-$~\cite{li2015coherent} and SiV$^-$~\cite{sipahigil2016integrated} centers. 
Compared to 1D PhC cavities, 2D PhC slab cavities can offer further improvements in the overall Purcell factor or cooperativity, as the 2D bandgap can inhibit unwanted SE channels out of the cavity through the reduction in the optical density of states~\cite{fujita2005simultaneous,englund2005controlling,noda2007spontaneous,lodahl2015interfacing}. However, due to the absence of high quality single-crystal diamond films, current fabrication methods of PhC slabs requires laborious thinning of bulk diamond films to wavelength-thickness films~\cite{faraon2012coupling,hausmann2013coupling,riedrich2014deterministic,Schroder2016_Review}. This fabrication process involves extensive reactive ion etching (RIE) processing along with complex micromanipulation, resulting in low yield of high quality devices~\cite{Schroder2016_Review}. Ion-slicing approaches~\cite{fairchild2008fabrication} present another alternative but entails crystal damage that can be mitigated only via overgrowth techniques~\cite{aharonovich2012homoepitaxial,gaathon2013planar,piracha2016scalable}.

Here, we demonstrate a process for fabricating 2D PhC slabs directly from bulk diamond. This technique, which extends our recently demonstrated fabrication of diamond 1D PhC nanocavities based on plasma quasi-isotropic undercutting of diamond~\cite{mouradian2017rectangular}, enables the complete release and suspension of PhC slabs on the bulk diamond surface. As with the 1D PhC cavity fabrication, which achieved record-high $Q$ factors near the NV ZPL, this new process also produces high-$Q$ cavities and fabrication consistency across a standard, bulk diamond chip.

As summarized in Figure 1, the fabrication process combines standard lithography with sidewall masking by atomic layer deposition (ALD) and quasi-isotropic dry etching. The process begins with RIE of bulk single-crystal diamond (3\,mm\,$\times$\,3\,mm\,$\times$\,0.3\,mm, {100} face, Type IIa CVD diamond with [N]$<$1ppm from Element Six) using an inductively coupled oxygen plasma (ICP\,=\,500W, RF\,=\,240W, T\,=\,32$^{\circ}$C, P\,=\,0.15\,Pa). This step uses a hard mask composed of a 230\,nm-thick silicon nitride (SiN) layer, produced by plasma-enhanced chemical vapor deposition, and patterned by electron-beam lithography (using ZEP-520A resist coated with Espacer conductive polymer, developed at 0$^{\circ}$C) and etched with tetrafluoromethane (CF$_4$) plasma (RF\,=\,200W). The optimized ICP-RIE parameters provide a 30:1 etch selectivity of diamond and near-vertical sidewalls. As sketched in Fig.\,\ref{fab}(a), this directional etch is $\sim$7$\times$ deeper than the desired PC slab thickness to facilitate release and precise tuning of the slab thickness in later steps.

Next, ALD of aluminum oxide (Al$_2$O$_3$) provides conformal coverage of the chip, including the sidewalls and PC holes [Fig.\,\ref{fab}(b)]. A CF$_4$ RIE step then selectively removes the top layer of Al$_2$O$_3$, which exposes the top facets of the chip [Fig.\,\ref{fab}(c)]. At this stage, the PhC region remains protected by the SiN mask and the sidewall by the Al$_2$O$_3$ coating. Diamond is exposed only in the bottom of etched patterns.

The diamond undercut relies on a quasi-isotropic oxygen plasma in an ICP-RIE chamber (ICP\,=\,900\,W, RF\,=\,0\,W, T\,=\,200$^{\circ}$C, P\,=\,3\,Pa), as illustrated in Fig.\,\ref{fab}(e). It is possible to image the diamond through the thin Al$_2$O$_3$ sidewall coatings to periodically check and control the extent of the undercut with high precision. We find that an optional directional ICP-RIE [Fig.\,\ref{fab}(e)]prior to the isotropic step reduces the total time to $\sim$8\,hours to fully suspend $\sim$4$\upmu$m-wide, $\sim$200\,nm-thick planar PhC slabs. Finally, a hydrofluoric acid wet etch removes the residual SiN and Al$_2$O$_3$ to reveal the air-clad diamond PhC devices [Fig.\,\ref{fab}(f)]. Figure\,\ref{fab}(g-i) shows various scanning electron micrographs of resulting diamond PhCs. 

\begin{figure}
\begin{center}
\includegraphics[width=3.5in]
{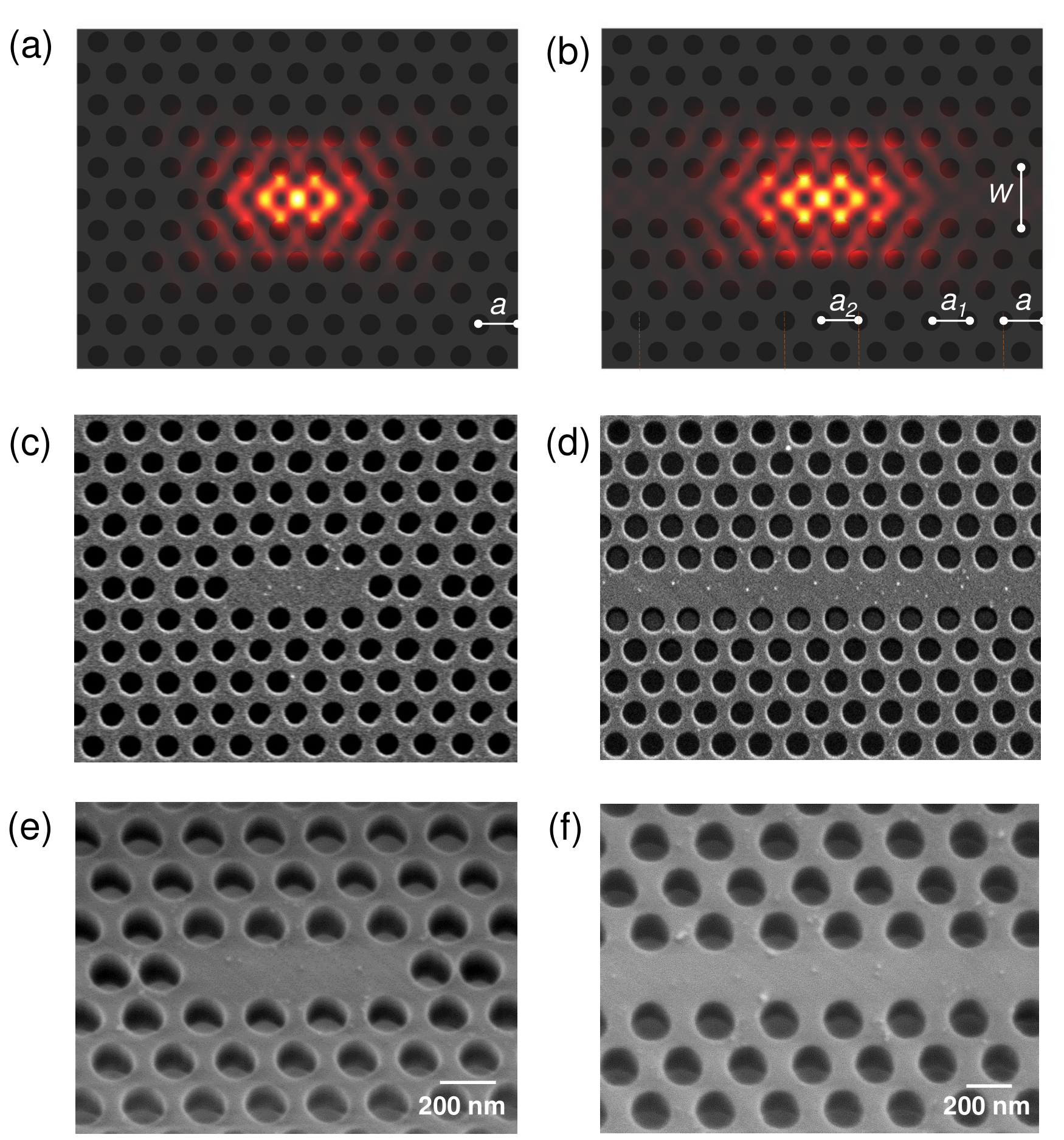}
\end{center}
\caption{
Normalized electric field density of (a) a modified $L3$-defect cavity with lattice constant $a = 214$\, nm and (b) a photonic heterostructure (HS) cavity with lattice constants $a = 210$\,nm, $a_1 = 1.025a$ and $a_2 = 1.05a$. (c,e) SEM images of fabricated $L3$ cavities. (d,f) SEM images of fabricated HS cavities.
}
\label{cavities}
\end{figure}

We fabricated two types of PhC cavity designs: $L3$-defect cavities and photonic heterostructure (HS) waveguide cavities~\cite{song2005ultra,bayn2008ultra} optimized for a fundamental mode at $\lambda = 637$\,nm. The former design consists of a line of three missing holes in an otherwise regular triangular-lattice PhC slab. The three adjacent holes on either side of the cavity region are slightly displaced outwards by $\{D1,D2,D3\} = \{0.219a,0.025a,0.2a\}$, where $a = 214$\,nm is the lattice constant, to minimize out-of-plane scattering. The PhC radius and slab height are $\{r,H\} = \{0.285a, a\}$. Figure\,\ref{cavities}(a) shows the electric field density of the fundamental TE-like cavity mode with quality factor $Q = 8,560$ and mode volume $V_m = 0.76 (\lambda/n)^3$, calculated using the Finite-Difference Time-Domain method (Lumerical Inc). The second design, shown in Fig.\,\ref{cavities}(b), is a line-defect PhC waveguide cavity with parameters $\{W,r,H\} = \{1.69a,0.275a, 0.96a\}$ with increasing lattice constants from $a = 210$\,nm to $a_1 = 1.025a$ and $a_2 = 1.05a$ at the center of the cavity. The HS cavity forms over four periods of $a_1$ and a single period of $a_2$, which maximizes the $Q/V$ ratio~\cite{bayn2008ultra}, where $Q = 250000$ and $V_m = 1.28 (\lambda/n)^3$ in simulation. Figures\,\ref{cavities}(c-f) show the corresponding scanning electron micrographs of the fabricated cavities. 

\begin{figure}[ht]
\begin{center}
\includegraphics[width=3.5in]
{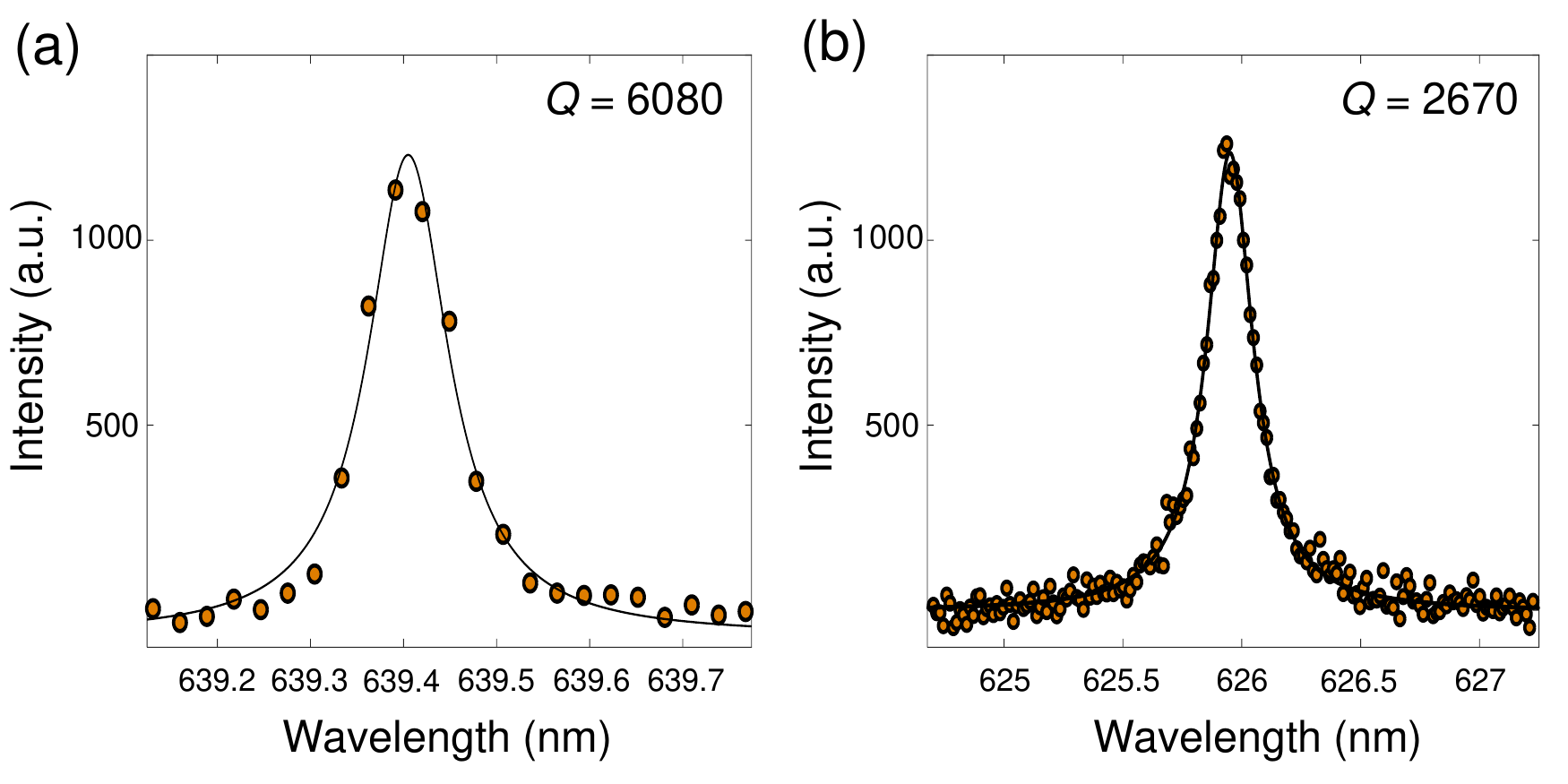}
\end{center}
\caption{
(a) $L3$ cavity resonance with $Q = 6080$ in the photoluminescence spectrum of NV$^-$ centers. Photonic HS cavity resonance with $Q = 2670$. 
}
\label{measurement}
\end{figure}

The fabricated cavities were characterized by photoluminescence (PL) confocal microscopy. Ensembles of native NV centers present in the diamond crystal serve as internal light sources for the cavities upon 532\,nm laser illumination. A spectrometer records the cavity-modified PL, from which we deduce the quality factors. These measurements reveal high-$Q$ resonances near the NV zero-phonon line for both nanocavity designs, with highest $Q_{L3} = 6080$ and $Q_{\text{HS}} = 2670$, shown in Fig.\,\ref{measurement}(a) and \ref{measurement}(b), respectively. The experimental $Q_{L3}$ is close to the simulated $Q = 8560$, which indicates that intrinsic cavity radiative losses limit the measured $Q$. However, the HS nanocavity $Q$ factor is two orders of magnitude lower than the simulated $Q$. This weak light confinement is likely due to mismatch in fabrication of this demanding design and the presence of slight thickness gradients; we believe with continued improvement in fabrication the $Q$ factor will likely increase.

The successful fabrication of high quality PhC slab nanocavities directly from bulk diamond significantly expands the potential for controlling light-matter interactions in diamond. Immediate experiments include NV-cavity resonant coupling, which is important to enhance the rates in photon-mediated quantum entanglement schemes. In these 2D PhC cavity modes, the electric field energy density is maximized in the the dielectric (as these modes are produced from the `air-band' of the Bloch vectors), which allows for excellent overlap between the emitter and the cavity mode. In addition, the PhC slab design allows the emitter to be relatively far from the etched surfaces, which improve the emitters' spectral stability~\cite{chu2014coherent,liu2017direct}. The PhC slab architecture greatly expands the types of devices that can be fabricated, including photonic crystal waveguides~\cite{lodahl2015interfacing,hood2016atom,lodahl2017chiral} that are of interest for investigating waveguide quantum electrodynamics and many-body physics with color centers. Other prospects that are now possible also include phononic and optomechanical crystals to extend the spin coherence of the SiV center~\cite{becker2017coherence}. Finally, the ability to fabricate planar PhC architectures on bulk dielectrics provides an excellent platform for compact nanophotonic device integration, and it should open up new possibilities in other fields, such as spectroscopy or classical light sources and modulators.

\begin{acknowledgments}
N.H.W was supported in part by the Army Research Laboratory Center for Distributed Quantum Information (CDQI) and Master Dynamic Limited. S.M. was supported by the NSF program ACQUIRE:``Scalable Quantum Communications with Error-Corrected Semiconductor Qubits.'' Fabrication was supported in part by the STC Center for Integrated Quantum Materials (CIQM), NSF Grant No. DMR-1231319. We thank James Daley, Mark Mondol, and Tim Savas at the Nanostrutures Laboratory at MIT for invaluable discussions and support during the development of this fabrication technique.
\end{acknowledgments}

\bibliography{references}
\end{document}